\documentstyle[epsfig,graphics]{article}
\begin{document}
\begin{center}
\huge {QCD-oriented nondiagonal GVDM}
\end{center}
\bigskip

\noindent E.V. Bugaev$^1$, B.V. Mangazeev$^2$, Y.V. Shlepin$^1$ \\

\noindent $^1$Institute for Nuclear Research,The Academy of Sciences  of
Russia, Moscow, Russia\\
\noindent $^2$Irkutsk State University, Irkutsk, Russia\\ 
\bigskip

{\bf \noindent Abstract.}
The nondiagonal generalized vector dominance model (GVDM) of photoabsorption
is elaborated using QCD-motivated picture of the 
$\gamma -\stackrel{\,~-}{qq}$-transition and subsequent meson dominated 
scattering of the $\stackrel{\,~-}{qq}$-pair on the nucleon.
The relativistic constituent quark model for a description of the meson
$\stackrel{\,~-}{qq}$-wave functions is used. The meson-nucleon scattering
is calculated in the two-gluon exchange approximation. It is shown that the 
destructive interference effects and corresponding cancellations in the 
photoabsorption cross section formula are small, so the GVDM predictions
are  incorrect  if  no  extra  cut-off  factors  in  GVDM  formulas  are
introduced.
\bigskip

\bigskip

{\bf \Large \noindent 1.Introduction.}
\smallskip

According to GVDM [1] the imaginary  part  of  the  transverse  forward
Compton scattering amplitude (or the transverse photon  absorption  cross
section) can be expressed in a form of the mass dispersion relation,
$$\sigma_T(Q^2,s)=\int
\frac{\rho_T(m^2,m'^2,s)m^2m'^2}{(m^2+Q^2)(m'^2+Q^2)}dm^2dm'^2. \eqno(1)$$
The spectral weight function $\rho_T$ is given by the  formula  of  GVDM
(in zero-width approximation):
$$\rho_T(m^2,m'^2,s)=\sum\limits_{n,n'}\delta(m^2-m^2_n)\delta(m'^2-m^2_
{n'})\frac {e}{f_n} \frac {e}{f_{n'}}\frac{ImT_{nn'}(s)}{s}. \eqno(2)$$
Here, $m_n$ is the vector meson mass, $f_n$  is  the  meson-photon  coupling
constant, $T_{nn'} (s)$ is an amplitude for  the  forward  meson-nucleon
scattering,
$$V_n+N\longrightarrow V_{n'}+N. \eqno(3)$$
The main problem, of course, is the description of  $\rho_T(m^2,m'^2,s)$
in     a     region     of      large      vector      meson      masses
$(m_n>>m_\rho,m_\omega,m_\varphi)$. A correct  treatment  of  the  heavy
masses would provide, in particular, the  convergence  of  the  integral
(1). The information about vector meson properties  in  the  heavy  mass
region is rather scarce, therefore, in pre-QCD era, for the proper choice
of the mass dependence of $\rho_T$ the motivation based on parton models was
used. It was shown [1] that, in general, $\rho_T$ can be chosen  in  the
form  compatible  with  Bjorken  scaling.  In  particular,  in  diagonal
approximation, when
$$\rho_T(m^2,m'^2,s)=\delta(m^2-m'^2)\rho_T(m^2,s), \eqno(4)$$
one needs, for this compatibility, the hadronic state continuum term  in
$\rho_T(m^2,s)$. If the $\rho_T$-function is  nondiagonal,  the  scaling
behavior of $Q^2\sigma_T$ is possible in  the  more  realistic  case  of
isolated hadronic states as  well.  If,  e.g.,  the  $\rho_T$-function
contains  large  negative  off-diagonal  contributions,  scaling  can  be
achieved through the destructive interference effects, i.e. through the
strong cancellations of diagonal and off-diagonal contributions  in  the
integral (1) (such a picture was confirmed by the direct calculation [2]
of $\rho_T(m^2,m'^2,s)$) in a framework of the covariant parton model).
The nondiagonal GVDM  [3-5]  based  on  this  picture  was  really  very
successful in a description of the nucleon structure functions at small
$Q^2$,  besides,  the  qualitatively  correct  $Q^2$-dependence  of  the
nuclear shadowing was obtained [6]. It was shown recently [7] that  this
model  even  predicts,  similarly  to  the  parton  model,   the   color
transparency effects. One should note, however, that the  choice  of  the
nondiagonal elements of $T_{nn'}$-matrix in this model  has  in  fact  no
connection with the  predictions  of  hadronic  models.  In  this  sense
nondiagonal GVDM uses some fictitious vector mesons. But the original GVDM's
idea is that a photon transforms virtually just  into
the genuine hadron  states  (those  observed  in  $e^+e^-$-annihilation)
which subsequently scatter from the target nucleon. Correspondingly, the
$f_n$-constants in Eq.(2) are expressed through the leptonic  widths  of
these states. The logic of GVDM should be such that the hadronic physics
is a starting point and the scaling behavior  of  $Q^2\sigma_T$  is  the
(nonnecessary, in principle) consequence.

In  last  few  years  many  works   appeared   applying   the   diagonal
approximation for $\rho_T$. Some of them introduce new  parametrizations
for $\rho_T(m^2,s)$-dependencies based solely on the experimental  data,
others sharply cut off the heavy mass tail of the  meson  mass  spectrum
adding,      instead,      the      large       softly       interacting
$\stackrel{~~ -}{qq}$-component [8] or the large direct component [9].

The common feature of all these recent works is, again, a  lack  of  the
attention paid to the purely hadronic aspects of the  problem.  Two  (at
least)  main  questions  should  be  studied:  i)  how   important   are
nondiagonal elements of $T_{nn'}$ in the integral (1) and  ii)  how  the
problem of the heavy meson masses is resolved (in particular, is there a
necessity in some extra cut-off factor in GVDM formulas).  

In  the  present  paper  we  try to answer  both  these  questions   using
relativistic constituent quark model of vector mesons and the  two-gluon
exchange approximation for a calculation of the meson-nucleon scattering
amplitudes.

A correct use of the hadronic basis is very important for a quantitative
description of the nuclear shadowing (which is quite sensitive just to a
space-time picture of the  process)  and,  especially, for  the  precise
calculation of $\sigma_{\gamma N}=\sigma_T(s,0)$ at  very  high  photon
energies  (at  the  energy  region  which  is  of  much   interest   for
astrophysics  and  cosmic  ray  physics  and  which  is   experimentally
inaccessible).

Concluding, one should add that now there are approaches [10,11] in  which
the hadronic basis in the photoabsorption description at small $Q^2$  is
completely discarded. The price for this is a necessity of an use of new
parameters  which  either have a rather dubious physical sense (as  the
virtuality dependent constituent quark mass [10,11]) or are  badly  known
(as the confinement radius [11]). 

\bigskip
{\bf \Large \noindent 2.The model of the hadronic amplitudes.}
\smallskip

We will use the simplest  model  of  VN-scattering:  two-gluon  exchange
approximation. For a calculation of the corresponding diagrams one  must
know, in particular, $\stackrel{~~ -}{qq}$-wave functions of the  mesons.
In a relativistic constituent  quark  model  these  wave  functions  are
obtained from the Bethe-Salpeter(BS) equation (we consider the  case  of
scalar identical quarks):
$$i(2\pi)^4\Phi(P,q)=\frac{1}{\Delta_1                     \Delta_2}\int
d^4q'K(q-q')\Phi(P,q'). \eqno(5)$$
Here,      $\Phi(P,q)$      is      the      BS      wave      function,
$P=k_1+k_2,~q=\frac{1}{2}(k_1-k_2),~k_{1,2}$   are   quark    4-momenta,
$\Delta_{1,2}=k^2_{1,2}-m^2_q,~K(q-q')$     is     the     $\stackrel{~~
-}{qq}$-interaction kernel.

For the utilization of  this  equation  it  is  convenient  to  use  the
quasipotential formalism in  a  light-front  form  (see,  e.g.,  [12,13]).
Variables needed  for  a  description  of  an  internal  motion  of  the
constituents  in  this  form  of  dynamics   are   light-front   momenta
$\stackrel{\to~~\to}{k_1,k_2}(\stackrel{\to~~~\to~~~~~~}{k=k_\bot,k_+})$
which transform covariantly under the kinematic Lorentz transformations.
Correspondingly,  $q_-$-component  of  the  relative   momentum   q   is
restricted.
The choice of a  concrete  form  of  this  restriction  is  not  unique,
however. The most simple and natural way is, in our opinion, an  use  of
the  covariant  condition  $Pq=0$  [13].  In  this  case  one  has  (if
$P_+P_-=M^2,P_\bot=0 ;~M$ is the bound state mass)     

$$q_-=-\frac{P_-q_+}{P_+}=\frac{-M^2q_+}{P^2_+};~~q^2=-q^2_\bot-\frac{M^2q
^2_+}{P^2_+}\equiv-q^2_\bot-M^2y^2.
\eqno(6)$$
Now   we    can    introduce    3-dimensional(3D)    "inner    momentum"
$\stackrel{\stackrel{\to}{\sim}}  {q}   (\stackrel{\sim~}{q_\bot}=q_\bot;
~\stackrel{\sim~}{q_3}\equiv My)$ which is,  according  to  Eqs.(6),  an
argument          of          the          interaction           kernel,
$K=K(\stackrel{\stackrel{\to~~~\to}{\sim~~~\sim}}{q-q'})$     .      The
corresponding reduction of the BS-equation can be performed [13]  by  the
integration both sides of Eq.(5) over $q_-$ and using the basic formula
$$\frac{1}{2}\int \frac {1}{\Delta_1  \Delta_2}dq_-=\frac{2\pi  i}{2P_+}
\frac {1}{M^2(\frac 14-y^2)-q^2_\bot-m^2_q}.\eqno(7)$$
The integration in Eq.(7) is done by contour methods and corresponds  to
putting one of the quarks on its mass shell.
The resulting 3D equation is
$$(\stackrel{\stackrel{\to}{\sim}}{q}^2+m^2_q-\frac
{M^2}{4})\Phi(\stackrel{\stackrel{\to}{\sim}}{q})=\frac{1}{16\pi^3M}\int d
\stackrel{\stackrel{\to}{\sim}}{q}
K(\stackrel{\stackrel{\to}{\sim}}{q}-\stackrel{\stackrel{\to}{\sim}}{q'})
\Phi(\stackrel{\stackrel{\to}{\sim}}{q'}).\eqno(8)$$
The                            wave                             function
$\Phi(\stackrel{\stackrel{\to}{\sim}}{q})(=\Phi(\stackrel{\to}{q_\bot},y))$
is simply connected with the Vqq-vertex function:
$$\Gamma(\stackrel{\stackrel{\to}{\sim}}{q})=\Phi(\stackrel{\stackrel{\to}{\sim}}{q})
(\stackrel{\stackrel{\to}{\sim}}{q}^2+m^2_q-\frac{M^2}{4}).\eqno(9)$$

Now we are able to calculate the two-gluon exchange  diagrams  starting
from the general 4D expressions and  reducing  them,  with  an  aid  of
Eqs.(7,9), to the 3D form. We keep only the terms of a leading order  in
s  and  neglect  the  longitudinal  momentum  transfer.  The   resulting
amplitude for the meson-meson scattering is
$$\begin{array}{l}
F(s,t=-\stackrel{\to}{Q}^2)=\displaystyle\frac{i}{(2\pi)^2}s \frac89 g^4 \int 
 \frac{d^2k_\bot}{(\frac{\stackrel{\to}{Q}}{2}-\stackrel{\to}{k_\bot})^2(\frac{\stackrel{\to}{Q}}{2}
+\stackrel{\to}{k_\bot})^2}\times \\ \qquad\\ \qquad
\times  [F_1(\stackrel{\to}  {Q}^2)-F_1(4\stackrel  {\to}   {k_\bot}^2)]
[F_2(\stackrel{\to}{Q}^2)-F_2(4\stackrel{\to}{k_\bot}^2)].\\
\end{array}
\eqno(10)$$  
Here, F is the meson formfactor given by the formula
$$F(Q^2)=\int                                              d^2q_\bot\int
dy(1-4y^2)\Phi(\stackrel{\to}{q_\bot},y)\Phi(\stackrel{\to}{q_\bot}+\frac{ 
\stackrel{\to}{Q}}{2},y).\eqno(11)$$
Similar (but not identical) expression for the  amplitude  was  obtained
long ago [14] employing the eikonal formalism.

The amplitude of the meson-nucleon scattering is obtained  from  Eq.(10)  by
the replacement:
$$[F_2(Q^2)-F_2(4\stackrel{\to}{k_\bot}^2)]\to                   \frac32
V(\stackrel{\to}{k_\bot},\stackrel{\to}{Q}).\eqno(12)$$
The V-factor describes the ggNN-vertex.We estimate  this  factor  using,
for simplicity, the approach of ref.[14]:
$$\begin{array}{l}
V(\stackrel{\to}{k_\bot},\stackrel{\to}{Q})\simeq\displaystyle \frac13 \sum\limits_i \langle
e^{i\stackrel{\to}{Q}\stackrel{\to}{x_i}}\rangle-\frac16\sum\limits_{i\ne j}
\langle e^{i[(\frac{\stackrel{\to}{Q}}{2}+\stackrel{\to}{k_\bot})\stackrel{\to}{x_i}+
(\frac{\stackrel{\to}{Q}}{2}-\stackrel{\to}{k_\bot})\stackrel{\to}{x_j}]}\rangle\simeq
\\ \qquad \displaystyle \simeq e^{-\frac{<r^2_N>}{6}\stackrel{\to}{Q}^2}-
e^{-\frac{<r^2_N>}{6}(\frac{\stackrel{\to}{Q}^2}{4}+3\stackrel{\to}{k_\bot}
^2)}
.
\end{array} \eqno(13)$$
Here, $<...>$ denotes the expectation value in the nucleon  bound  state
and it  is  assumed  that  the  three  quarks  have  the  same  Gaussian
distribution; $<r^2_N>$ is the mean squared radius of the nucleon.

For the solution of Eq.(8) we assume that kernel K  has  only  the  long
range confining term of the hadronic oscillatior type [13]: 
$$K(\stackrel{\stackrel{\to}{\sim}}{q}-\stackrel{\stackrel{\to}{\sim}}{q'})
=(2\pi)^3\omega^2_{\stackrel{\,~-}{qq}}(\stackrel{\to 2}{\bigtriangledown_
{\stackrel{\sim}{q}}}+\omega^{-2}_0)\delta^3(\stackrel{\stackrel{\to}{\sim}}{q}-\stackrel{\stackrel{\to}{\sim}}{q'})
, \eqno(14)$$
with two  parameters: $\omega^2_{\stackrel{\,~-}{qq}}$  (a "spring   constant")   and
a  zero-point   energy
$\omega_0$.
Equation (8) with this kernel formally coincides with the equation for a
quantum-mechanical 3D-oscillator:
$$(\stackrel{\stackrel{\to}{\sim}}{q}^2+m^2_q-\frac{M^2}{4})\Phi(\stackrel{\stackrel{\to}{\sim}}{q})
=\frac{\omega^2_{\stackrel{~-}{qq}}}{2M}(\stackrel{\to 2}{\bigtriangledown_
{\stackrel{\sim}{q}}}+\omega^{-2}_0)\Phi(\stackrel{\stackrel{\to}{\sim}}{q})
. \eqno(15)$$
Solutions of Eq.(15), its eigenfunctions and eigenvalues are well  known.
We will use them for a description of the $\rho$-family. The mass spectrum
of radial excitations is:
$$\frac{1}{2\beta^2}(\frac{M^2}{4}-m^2_q+\beta^4\omega^2_0)=N+\frac32~
;~~N=0,2,4,...~. \eqno(16)$$
Deriving Eq.(16) we assume that $\beta^2\equiv \omega_q/\sqrt{2M}$ is a constant (i.e., is  independent  on
M). In this case the meson mass spectrum has the form $m^2_n=a+bn$.
With the numerical values
$$m_\rho=0.77,~~m_{\rho'}=1.45,~~m_q=0.3 $$
one has
$$\beta^2=0.094 ~Gev^2~,~~\omega^2_0=0.04~Gev^2$$
and, finally,
$$m^2_n\cong m^2_\rho(1+2.55n)~;~~n=0,1,2,...~. \eqno(17)$$
The $\rho$-meson wave function is
$$\Phi_0(\stackrel{\stackrel{\to}{\sim}}{q})=\Phi_0(
\stackrel{\to~}{q_\bot},y)=N_0
exp[-(q^2_\bot+m^2_\rho y^2)/2\beta^2], \eqno(18)$$
and $N_0$ is determined from Eq.(11), using the condition F(0)=1. In the
$(\stackrel{\to~}{r_\bot},y)$-representation one has
$$\Phi_0(\stackrel{\to~}{r_\bot},y)=N_0 exp[-r^2_\bot\beta^2/2]
exp[-m^2_\rho y^2/2\beta^2]. \eqno(19)$$
Using  the  $(\stackrel{\to~}{r_\bot},y)$-space,   the   $V_nN$-scattering
amplitude can be written as
$$F_{nn}(s,t)=\int                     d^2r_\bot                      dy
F_{\stackrel{\to~}{r_\bot}}(s,t)\Phi^2_n(\stackrel{\to~}{r_\bot},y)\equiv
<n\mid F_{\stackrel{\to~}{r_\bot}}(s,t)\mid n>, \eqno(20a)$$
$$F_{\stackrel{\to~}{r_\bot}}(s,t)=i\frac{16}{3}\alpha^2_ss\int
\frac{d^2k_\bot
V(\stackrel{\to~}{k_\bot},\stackrel{\to}{Q})}{(\frac{\stackrel{\to}{Q}}{
2}-\stackrel{\to~}{k_\bot})^2(\frac{\stackrel{\to}{Q}}{
2}+\stackrel{\to~}{k_\bot})^2
}\{ e^{-i\frac{\stackrel{\to}{Q}}{2} \stackrel{\to~}{r_\bot}}-e^{-i
\stackrel{\to~\to~}{k_\bot r_\bot}}\}. \eqno(20b)$$
Here,  $F_{\stackrel{\to~}{r_\bot}}$  is  an  "eigenamplitude",  i.e.  an
amplitude for the scattering of the $\stackrel{~~  -}{qq}$-pair  with  a
fixed $\stackrel{\to~}{r_\bot}$ on the nucleon. Eq.(20a) could be written
without any derivation, using only the simple physical fact that, due to
the large lifetime of $\stackrel{~~ -}{qq}$-fluctuations at large s  the
values $\stackrel{\to~}{r_\bot},y$ are "frozen" in the scattering process
(and therefore they are "eigenvalues" of  the  scattering  matrix).  The
concrete expression for $F_{\stackrel{\to~}{r_\bot}}$ (Eq.(20b)) is given
by the model.

Integrating over asimutal angles in Eqs.(20) we reduce the problem to  a
calculation of $F_{\stackrel{\to~}{\mid r_\bot\mid}}\equiv  F_{r_\bot}$.
This amplitude depends only on two parameters, $\alpha_s$  and  $\mu_g$,
effective gluon mass (omitted in the above-cited  expressions  for  brevity's
sake). Going into  impact  parameter  space,  we  introduce  the  opaque
function 
$$\Omega_{r_\bot}(s,b)=\frac{1}{2\pi     i}\int      \frac      {1}{4\pi
s}F_{r_\bot}(s,t)~e^{\stackrel{~~\to\to}{iQb}}d^2Q. \eqno(21)$$
The  numerical  calculation  shows   that   $\Omega_{r_\bot}$   can   be
parametrized with a large accuracy by the Regge-type expression:
$$\Omega_{r_\bot}(s,b)=\frac{\sigma(r_\bot)}{4\pi
B_{r_\bot}}exp(-\frac{b^2}{2B_{r_\bot}}), \eqno(22)$$
where 
$$\sigma(r_\bot)=\frac{1}{s}ImF_{r_\bot}(s,0);~~B_{r_\bot}=\frac{\sigma(r
_\bot)}{4\pi\Omega_{r_\bot}(s,0)}. \eqno(23) $$   
We took in this analysis $\mu_g=\mu_\pi$ and normalized $\sigma(r_\bot)$ on 
the pion data at medium energies $(\sqrt{s}=10~Gev)$, in accordance with 
the additive quark model relation
$$\sigma_{\rho p}=\frac12(\sigma_{\pi^+ p}+\sigma_{\pi^- p}). \eqno(24)$$    
 For this normalization we used the unitarized scattering amplitude
$$T_{\rho \rho}(s,0)=\int <\rho\mid 1-e^{-\Omega_{r_\bot}(s,b)}\mid\rho>d^2b.
\eqno(25)$$

Up to now in our model $F_{r_\bot}\sim s$ so that $\Omega_{r_\bot}$ does not 
depend on the energy. To take into account this dependence we modify Eq.(22)
adding a new Regge-type term:
$$\Omega_{r_\bot}(s,b)=\frac{\sigma(r_\bot)}{4\pi} \left\{\frac{1}{B_{r_\bot}}
e^{-\frac{b^2}{2B_{r_\bot}}}+\frac1R \frac{1}{B_{r_\bot}+2\alpha'_F \xi}
e^{\Delta_F \xi}e^{-{\frac{b^2}{2(B_{r_\bot}+2\alpha'_F \xi)}}}\right\}, \eqno(26)$$
$$\xi=ln\frac {s}{s_0}-\frac{i\pi}{2};~~\Delta_F=\alpha_F-1>0;~~\alpha'_F\ne0.$$
Writing Eq.(26) we suppose a two-pole Regge-parametrization [15] of the opaque 
function; by assumption, both trajectories give at small $\xi$ the same 
diffraction slopes and R does not depend on $r_\bot$. Three new parameters
$(R,\Delta_F,\alpha'_F)$ are determined from data on a s-dependence of hadronic 
total amplitudes.

Now we have all necessary for a calculation of the amplitudes. It is evident that 
 the nondiagonal amplitudes are given by a simple generalization of Eq.(20a):
$$F_{nn'}(s,t)=<n\mid F_{\stackrel{\to~}{r_\bot}}(s,t)\mid n'>. \eqno(27)$$
In the nondiagonal case one has, instead of the elastic scattering $V_nN\to V_nN$,
 the diffraction dissociation $V_nN\to V_{n'}N$.

\bigskip
{\bf \Large \noindent 3.Cut-off factors.}
\smallskip  

The first stage of the photoabsorption process is the $\gamma\to 
\stackrel{~~-}{qq}$-transition. The  differential  probability  of  this
transition                                                           is
$$dP_{\stackrel{\,~-}{qq}}=C\frac{1}{\mu^2_\bot}\{x^2+(1-x)^2+\frac{2m^2_q}
{\mu^2_\bot}x(1-x)\} dxdp^2_\bot. \eqno(28)$$
Here, C is a known constant, $\mu_\bot$ is the quark transverse mass, 
$\mu_\bot=\sqrt{p_\bot^2+m^2_q}$, x is the fraction of the photon  3-momentum
carried by the quark. An invariant mass of the $\stackrel{~~-}{qq}$-pair
is  $$M^2_{\stackrel{\,~-}{qq}}=\mu^2_\bot/x(1-x).  \eqno(29)$$  The  last
term  in  the  curly  brackets  in  Eq.(30)  is  of  the   order   $\sim
m^2_q/M^2_{\stackrel{\,~-}{qq}}$ and can  be  safely  neglected  (in  this
section  $m_q$  is  the   current   quark   mass).   Going   over   from
$(p^2_\bot,x)$-
 to $(M^2_{\stackrel{~-}{qq}},x)$-variables in Eq.(28) we obtain
$$dP_{\stackrel{\,~-}{qq}}\cong C\frac{1}{M^2_{\stackrel{\,~-}{qq}}}[x^2+(1-x)^2]
dxdM^2_{\stackrel{\,~-}{qq}}. \eqno(30)$$

It  is  easy  to  show  that  the  average  transverse   size   of   the
$\stackrel{~~-}{qq}$    -pair    is     given     by     the     formula
$$\stackrel{-~~}{r_\bot}=\upsilon_{\bot,relative}\cdot \tau_{fe}\cong
\frac{p_\bot}{\mu^2_\bot}(1+\frac{Q^2}{M^2_{\stackrel{~-}{qq}}})^{-1}
~\stackrel{m_q=Q^2=0}{\longrightarrow}~\frac1p_\bot. \eqno(31) $$
 In the present paper we consider only the case $Q^2=0$. Evidently,
$\stackrel{-~~~~~}{{r^{max}_\bot}}(\sim m^{-1}_q)$ is much larger than  a
typical  transverse  size  of  hadrons.  Our  basic  assumption  is  the
following: an interaction  of  the  $\stackrel{~~-}{qq}$-pair  with  the
nucleon is meson-dominated if (and only if) this  pair  is  wide  enough
(i.e. if  $\stackrel{-~~}{r_\bot}>r^0_\bot\sim\sqrt{<r^2_h>}$;  only  in
this case confinement forces are effective and pull the pair's particles
together). It is known that the narrow  pairs  weakly  interact  with  a
nucleon (due to the  color  transparency  phenomenon).  Therefore,  the
corresponding nonVDM contribution is, fortunately,  small  and  can  be
taken into account by a slight  change of the $r^0_\bot$-parameter. Very
narrow pairs (with $p_\bot \ge 2~Gev$) interact with  a  nucleon  purely
pertubatively and  must  be  considered  separately  (the  corresponding
"anomalous" contribution to $\sigma_{\gamma N}$  is  essential  only  at
very high energies).

The  restriction  of the $\stackrel{~~-}{qq}$-pair's   phase   volume   was
discussed by many authors beginning from its suggestion in  ref.[16].  It
follows from Eqs.(29,30) that, at fixed  $M^2_{\stackrel{\,~-}{qq}}$,  the
relative part of pair's phase  volume  having  $p_\bot$  in  the  limits
$(\sim m_q\div p_\bot^{max})$ is given by 
$$\eta                                                           \approx
3\left(\frac{p_\bot^{max}}{M_{\stackrel{~-}{qq}}}\right)^2,~~for~
M^2_{\stackrel{~-}{qq}}\gg(p_\bot^{max})^2. \eqno(32)$$
Using  Eq.(31)  we   introduce   the   restricting   factors.Identifying
$M_{\stackrel{~-}{qq}}$ with $m_n$, we have for each vector meson:
$$p_{\bot             n}^{max}=(\alpha\sqrt{<r^2_n>})^{-1};~~\eta_n\cong
3\left(\frac{p^{max}_{\bot n}}{m_n}\right)^2. \eqno(33)$$
We assume that parameter $\alpha$ is the same for all mesons. Evidently,
the restriction is absent if  $M_{\stackrel{~-}{qq}}<2p_\bot^{max}$  (it
appears that this inequality is valid for $\rho,\omega$-mesons only).

Finally, in GVDM formulas the following substitutions must be done:
$$\frac   ef_n   \longrightarrow   \frac   ef_n\sqrt\eta_n\equiv   \frac
e{\stackrel{\sim}{f_n}}. \eqno(34)$$

The mean square transverse radii of  vector  mesons  can  be  calculated
using the wave functions described in the previous section.

\bigskip
{\bf \Large \noindent 4.Results and conclusion.}
\smallskip

Our formula for the photoabsorption cross section is $(Q^2=0)$
$$\sigma_{\gamma
N}=\sum\limits_{n,n'}\frac{e^2}{\stackrel{~\sim~\sim~}{f_n
f_n'}}\frac{ImT_{nn'}}{s}\equiv\sum\limits_{n,n'}\sigma^{nn'}_{\gamma N}.
\eqno(35)$$
The coupling constants $f_n$ are simply connected with the lepton widths
$\Gamma_n(V_n\to  e^+  e^-)$.  In  principle,  these  widths  should  be
calculated with an aid of the quasipotential formalism  used  above.  It
will be done in  a  separate  paper.  Now  we  assume,  as  usual,  that
$\Gamma_n\sim m_n^{-1}$ (here n is, as earlier, the meson number in  the
family). From this it follows that $f_n\sim m_n$.

In the table 1 we present our results for the $\rho$-family contribution
to $\sigma_{\gamma N}~(\sqrt{s}=10Gev)$. Each number in the second  line
of the table is one of the  terms  in  the  sum  of  Eq.(35)  calculated
without cut-off factors. The basic  coupling  constant  is  well  known,
$f^2_\rho /4\pi=2.25$. 
The third line of the table contains $\sigma^{nn'}_{\gamma  N}$  obtained
after    an    insertion    of    the    cut-off    factors,     $f_n\to
\stackrel{\sim~}{f_n}$. 

\smallskip
\smallskip
\begin{tabular}{|c|c|c|c|c|c|c|}
\hline
 & $\rho\rho$  &   $\rho'\rho'$   &   $\rho''\rho''$   &   $\rho\rho'$    &
$\rho\rho''$ & $\rho'\rho''$  \\
\hline
$\sigma_{\gamma N}^{nn'}$ without cut-off factors & 77.9 & 41.86 & 28.49 & 12.9 & -5.6 & -5.04\\
\hline
$\sigma_{\gamma N}^{nn'}$ with cut-off factors & 77.9 & 10.56 & 2.66 & 6.47 & -1.71 & -0.77\\
\hline
\end{tabular}

\smallskip
\smallskip

 Parameter  $\alpha(=0.474)$  was  found  by  the
comparison of the theoretical $\sigma_{\gamma N }$ with the experimental
value $(\sim 115~ \mu bn)$. The contribution  of  $(\omega,\phi)$-families
was estimated in diagonal approximation with the result:
$$\sum\limits_{n(\omega,\phi)}\sigma^{nn}_{\gamma
N}=8.65+\frac{2.3}{\alpha^2}. \eqno(36)$$
The energy dependence of $\sigma_T$ is shown  on  fig.1.  The  following
values of the parameters were used:
$$R=22;~~\Delta_F=0.25;~~\alpha_F=0.13.$$

 The numerical results  obtained  in  the  present  model  lead  to  the
following conclusions.

1. If  no  cut-offs  are  introduced,  GVDM  is  not  able  to  describe
photoabsorption data. Even the simplest variant of the  GVDM  containing
only $\rho$ and $\rho'$ give too large  value  of  $\sigma_{\gamma  N}$.
Nondiagonal  contributions  are  not   negligibly   small.   Destructive
interference effects proposed in [3-5] are not effective (in particular,
the largest nondiagonal term $(\rho \rho')$ is positive). 

2. The introduction of the cut-off factors motivated by QCD can give  the
correct predictions. This is reached without  nonnatural  break  of  the
meson mass spectrum (the value of the heaviest mass in GVDM  expressions
is determined solely by the condition that the longitudinal size of  the
fluctuation must exceed the target size). In the present model only  one
parameter $(\alpha)$ is needed  for  the  description  of  all  cut-offs
(despite  the  fact  that  $p^{max}_{\bot}$  values  are  different  for
different mesons). In the  scheme  with  the  cut-offs  the  nondiagonal
contributions have the same order of magnitude as neighbouring  diagonal
ones, so GVDM developed in the present paper is an essentially nondiagonal model.

\end{document}